The metal-insulator transition and its relation to magnetic structure in $(LaMnO_3)_{2n}/(SrMnO_3)_n$ superlattices

A. Bhattacharya[1,2] *, S. J. May[1], S. G. E. te Velthuis[1], M. Warusawithana[4], X. Zhai[4], A. B. Shah[5], J.-M. Zuo[5], M. R. Fitzsimmons[6], S. D. Bader[1,2], J. N. Eckstein[4].

[1] Materials Science Division and [2] Center for Nanoscale Materials, Argonne National Laboratory, Argonne, IL 60439.

[4] Department of Physics and [5] Department of Materials Science and Engineering, University of Illinois at Urbana-Champaign, IL 60801.

[6] Los Alamos National Laboratory, Los Alamos NM 87545.

Abstract:

Superlattices of $(LaMnO_3)_{2n}/(SrMnO_3)_n$ ($1 \leq n \leq 5$), composed of the insulators $LaMnO_3$ and $SrMnO_3$, undergo a metal-insulator transition as a function of $n$, being metallic for $n \leq 2$ and insulating for $n \geq 3$. Measurements of transport, magnetization and polarized neutron reflectivity reveal that the ferromagnetism is relatively uniform in the metallic state, and is strongly modulated in the insulating state, being high in $LaMnO_3$ and suppressed in $SrMnO_3$. The modulation is consistent with a Mott transition driven by the proximity between the $(LaMnO_3)/(SrMnO_3)$ interfaces. Disorder localizes states at the Fermi level at the interfaces for $n \geq 3$. We suggest that this disorder is due to magnetic frustration at the interfaces.

An interface between two strongly correlated materials creates a distinct environment for allowing new collective states to emerge. As an example, a metallic state is obtained at an interface between the band-insulator $SrTiO_3$ and the Mott-insulator $LaTiO_3$.[1] This happens as electrons from the $Ti^{3+}$ states in $LaTiO_3$ leak into the $Ti^{4+}$ states in $SrTiO_3$, and a non-zero density of states (DOS) develops at the Fermi level ($E_F$) near the interface. In the $Mn^{3+}/Mn^{4+}$ mixed-valence manganites, diverse phases with charge, magnetic and structural order emerge as a function of doping, including a double-exchange mediated ferromagnetic (FM) metal.[2] The end-members of their phase diagrams are antiferromagnetic insulators (AFI). In the $La_{1-x}Sr_xMnO_3$ system, $LaMnO_3$ (LMO) with nominally $Mn^{3+}$ $t_{2g}^3 e_g^1$ occupancy, is an AFI with strong Mott-Hubbard/charge-transfer Coulomb correlations in a half-filled $e_g$ band.[3] The $Mn^{3+}$ sites are also Jahn-Teller active, and this leads to an $A$-type orbital-ordered layered antiferromagnet (FM in plane, AF between planes) in bulk samples at low temperatures. At the other end, $SrMnO_3$ (SMO), is a band-insulator with $G$-type or cubic AF order in the high-spin $Mn^{4+}$ $t_{2g}^3 e_g^0$ configuration. In the usual bulk samples of $La_{1-x}Sr_xMnO_3$, the $A$-site is populated randomly by $La^{3+}$ and $Sr^{2+}$ cations, allowing the Mn cation on the $B$-site to have a mixed-valence of $Mn^{3+}/Mn^{4+}$. With current state-of-the-art techniques,[4] atomically sharp SMO/LMO interfaces can be created where charge leakage may lead to mixed-valence states at coherent two-dimensional (2-D) layers. The correlations between the spin, charge and orbital degrees of freedom at the interface may give rise to states with properties that are distinct from LMO, SMO or any part of the randomly alloyed phase diagram.

In a simplified model of LMO/SMO interfaces by Lin $et\ al.$, the layers of $La^{3+}$ are considered to be donor ions in a lattice of $SrMnO_3$,[5] with the doped electrons going into the $e_g$ states on the Mn sites, and bound to the $La^{3+}$ layer via an attractive Coulomb potential. The hopping amplitude $t$ between neighboring $Mn^{3+}/Mn^{4+}$ sites at the LMO/SMO interface, which depends on the alignment between the Mn spins via double-exchange, causes the electronic profile at the interface to be intrinsically smeared. The Thomas-Fermi length-scale for charge leakage is given by $L_{TF} = a(\frac{t}{e^2/\varepsilon a})$, where $\varepsilon$ is the dielectric function and $a$ is the lattice constant, and is the ratio between $t$ and the attractive Coulomb interaction. This is calculated to be 1-3 unit cells (u.c.), depending upon the values of $\varepsilon$ and $t$.[5] In analogy with the titanates,[6] this interface could also have a finite DOS at $E_F$, and may be FM for appropriate conditions, in which case a spin-polarized 2-D electron gas could be realized. It is known that in short-period superlattices (*e.g.* $(LMO)_{2n}/(SMO)_n$ for $n$=1) where the LMO/SMO interfaces are brought into close proximity, the transport and magnetic properties are quite similar to those of the corresponding random alloy.[7] At the simplest level, it is useful to think about this in terms of a Mott-transition[8], in which the attractive Coulomb potential of the $La^{3+}$ layers at an LMO/SMO interface is screened by electrons from the neighboring interface. Below a critical separation, the attractive potential becomes too weak to provide a bound state, causing the electrons to become delocalized between the interfaces. This gives rise to a homogeneous 3D charge distribution similar to the random alloy, but without the *A*-site disorder.[9,10] Upon increasing the separation between interfaces, the electrons eventually become bound near the $La^{3+}$ ions, and relatively strong modulation of electron density and the concomitant order parameters is expected within the superlattice.[5] Transport and magnetization measurements have been used to study similar structures, and there has also been a recent study of the electronic reconstruction at these interfaces.[11] However, there has not yet been a direct measurement of the underlying structure of the relevant order parameters at these buried interfaces that would allow us to correlate the transport and magnetic properties to them.

In this Letter, we investigate highly-ordered digital superlattices of $(LaMnO_3)_{2n}/(SrMnO_3)_n$, where the overall stoichiometry is equivalent to the FM metal $La_{0.67}Sr_{0.33}MnO_3$. These superlattices undergo a metal-insulator transition as a function of $n$ at low temperatures ($T$). For $n \leq 2$, a FM metal is obtained, with relatively uniform ferromagnetism, closely related in properties to those of the random alloy $La_{0.67}Sr_{0.33}MnO_3$. For $n \geq 3$, the superlattices are insulating state at the lowest $T$. Using neutron reflectivity, we show that in the insulating state for $n \geq 3$, the ferromagnetism is strongly modulated with a periodicity commensurate with the superlattice, with suppression of ferromagnetism in the $SrMnO_3$. This is evidence for the modulation in properties characteristic of a Mott-like transition driven by the proximity between LMO/SMO interfaces. Furthermore, the localized states that emerge for $n \geq 3$ are not like the gapped states observed in LMO or SMO, but rather are states near $E_F$ that have been localized by disorder. We suggest that these localized states are interfacial in nature, and that the disorder that localizes them may have an intrinsic magnetic origin due to frustrated spins between adjacent AF and FM layers within these structures.

Superlattices of $[(LMO)_{2n}/(SMO)_n]_p$ were grown on strontium titanate (STO) substrates etched in dil HCl, with ozone-assisted molecular beam epitaxy (MBE). In all our films, $n$ integer unit-cells of SMO, were followed by $2n$ integer unit-cells of LMO, repeated $p$ times, where $p \bullet (n+2n)$ is an integer close to 80. This strategy, referred to as 'digital synthesis', has recently been used in the context of both manganites[9,12] and titanates.[13] The superlattices are fully strained to the substrate for the thicknesses presented in this work (~ 300 Å), leading to compressive strains of ~2 % for LaMnO$_3$ and a biaxial tensile strain of ~2.6% for SrMnO$_3$ at 300K and lower.[14,15,16] Scanning transmission electron microscopy (STEM) of an $n$ = 4 sample revealed well-defined interfaces between the SrMnO$_3$ and LaMnO$_3$ layers [Fig. 1(b)].[17] Grazing and high angle x-ray scattering confirm that the structures have smooth interfaces over macroscopic distances [Fig. 1(a)].

The resistivity ($\rho$) vs. temperature ($T$) is shown in Fig. 2 for a series of digital superlattices from $n$ =1 up to $n$=5, and compared to a random alloy film of identical composition La$_{0.67}$Sr$_{0.33}$MnO$_3$. The $\rho(T)$ increases by over eight orders of magnitude with increasing $n$ at the lowest $T$. Near the nominal Curie temperature $T_C$, all superlattices show a drop in $\rho$, consistent with double-exchange mediated ferromagnetism (for $n$=5, $T_C$ is ~40 K less than the temperature for the downturn in $\rho$). The $n$ =1 superlattice has a lower $\rho$ at low $T$ than the random alloy, probably a result of lower impurity scattering of charge carriers. At any given $T$, $\rho$ increases upon increasing $n$, and an insulator is obtained at the lowest $T$ for $n \geq 3$. A 20 u.c. film of LMO grown on STO under identical conditions was insulating and obeyed $\rho = \rho_0 \exp(E_A / kT)$ with $E_A$ = 125 meV above the magnetic ordering temperature of 150 K, where there is an inflection. This film was found to be FM with a saturation moment $M_S$ = 3.25 $\mu_B$/Mn. The FM behavior in these very thin films[18] may be due to strain and/or a deficiency of La. Superlattices of (SMO)$_3$/(LMO)$_1$ (corresponding to $x$=0.75) were also found to be strongly insulating (Fig. 2, inset)[19], and no signatures of magnetic ordering could be detected with a SQUID (Superconducting Quantum Interference Device) magnetometer. The $T_C$ value for the random-alloy is 355 K and the saturation magnetization $M_S$ = 3.22 $\mu_B$/Mn, while for the $n$ = 1 superlattice, $T_C$ = 340 K, and $M_S$ = 3.0 $\mu_B$/Mn. The coercivity $H_c$ is nearly identical between the two samples at 10 K. For the $n$=1 sample, if we assume a maximum possible FM moment of 4 $\mu_B$/Mn in the LMO layers, we can set a lower bound on the FM moment in the SMO layers of 1 $\mu_B$/Mn, in order for $M_S$ to be equal to the measured value. However, such a distribution of magnetization would give rise to strongly insulating properties. The similarity of the $n = 1$ and random-alloy in $\rho(T)$ and magnetic properties point to a more homogeneous electronic and magnetic structure within the $n = 1$ sample.

As $n$ increases, the value of $H_c$ increases, and that of $M_S$ is strongly suppressed for $n \geq 3$ [Fig. 3]. In order to probe the underlying magnetic order, we have used polarized neutron reflectometry (PNR) to measure the magnetic structure for wave-vectors perpendicular to the plane ( $q_\perp$ ) in an $n = 5$ and $n = 3$ superlattice. The measurements were carried out using the polarized neutron reflectometer ASTERIX[20] at the Los Alamos Neutron Scattering Center of Los Alamos National Laboratory. Figure 4 (a) shows the PNR data at 300 K in a field of 5.5 kOe. This temperature is well above $T_C$, and the reflectivity for the two spin states ($R^+$ and $R^-$, incident neutrons polarized parallel and opposite to the field, respectively) is the same. Due to the similarity in the nuclear scattering length for La and Sr, the structural Bragg peak visible in the x-ray reflectivity at $q_\perp$ = 0.108 Å$^{-1}$ (see inset) is barely visible in the case of neutrons. After field cooling in 5.5 kOe to 10 K [Fig 4(b)], there is a significant difference between $R^+$ and $R^-$, evidence for ferromagnetism in the sample. Furthermore, a strong Bragg peak emerges at $q_\perp$ = 0.103 Å$^{-1}$, and a second order peak is visible near $q_\perp$ = 0.2 Å$^{-1}$. According to our best fit to the PNR data,[21] the magnetization is strongly modulated commensurate with the superlattice period, with a maximum near 3.8 $\mu_B$/Mn within the LaMnO$_3$, and a minimum of less than 0.1 $\mu_B$/Mn in the SrMnO$_3$. This rules out all scenarios that would allow a reduced moment with a uniform distribution (*e.g.* a uniformly canted AF state), or even those where most of the modulation occurred in-plane. The low moment in some regions of the LMO (< 2.6 $\mu_B$/Mn ) suggests a canted state. The integrated magnetization in the superlattice extracted from PNR is found to be within 6% of the value measured with the SQUID magnetometer. For $n = 3$ at $T = 10$ K, we also observe a magnetic Bragg peak at $q \sim 0.175$ Å$^{-1}$, corresponding to a modulated magnetization. The calculated reflectivities that best reproduce the data suggest an amplitude for the modulation that is lower in magnitude than in the $n = 5$ sample.

Since the length-scale for charge leakage from calculations is 1-3 u.c., we would expect that for $n > 3$, the LMO/SMO interfaces are relatively isolated from one another, and the in-plane transport properties of the superlattices are closely related to those for a single interface. We note that $\rho(T<30$ K) for the most insulating superlattice is significantly lower than the LMO film at 100 K, and also the (SMO)$_3$/(LMO)$_1$ superlattice. The $\rho$ is also three orders of magnitude lower than known values for SMO films[7] on STO at 400 K. Thus, the high $n$ superlattices are *not* like either SMO or LMO. We argue that this lower resistivity is due to carriers near the LMO/SMO interface, where the insulators are effectively 'doped' by the diffusion of charge carriers between LMO and SMO to create a distinct conducting state. We find that for $n \geq 3$, $\rho$ for $T = 2 - 28$ K is consistent with Mott's variable range hopping (VRH) in 3D *i.e.* $\rho = \rho_o \exp(T_O/T)^{1/4}$ [Fig. 5(a)]. In this scenario, there is indeed a non-zero DOS at $E_F$, but these states are localized due to disorder, and their exponential tails decay over a characteristic localization length $\xi_L$. The charge carriers move between local sites via thermally assisted hops, preferentially to sites close in energy, over distances $L_{hop} \sim \xi_L (T_0/T)^{1/4}$. From the values of $T_0$ obtained from fitting to the data, the localization length can be calculated from the expression $\xi_L = \left\{[18/k_B T_0 N(E_F)]\right\}^{1/3}$, where $N(E_F)$ is the DOS at $E_F$.[22] For the $n$=3 sample which lies near the border of the metal-insulator transition, $\xi_L$ = 180Å. For the $n$=4 superlattice, $\xi_L$ = 11.4 Å, and $L_{hop}$ (10 K) = ~ 18.5 Å at 10 K, while for the $n$=5, $\xi_L$ = 3.6 Å, and $L_{hop}$ ~ 14 Å at 10 K. For $n$ = 4 and 5, the $\xi_L$ is smaller than the distance between the two nearest LMO/SMO interfaces. The calculated hopping lengths at 10 K are comparable to the distance between the nearest interfaces, and become larger at lower temperatures. The hopping exponent is consistent with 3D transport. This suggests that the carriers are localized in regions of extent $\xi_L$ near a single interface but can hop between states at different interfaces.

The PNR data indicate that the high $n$ superlattices have FM regions next to non-FM regions. If the non-FM regions are AF, then we expect to observe the consequences of competing AF/FM interactions with magnetic pinning, frustration and canted order. We have measured the magnetoresistance (MR), defined as *[R(H)-R(0)]/R(0)*, where $H$ is the magnetic field. The insulating superlattices have MR that is seven times or greater in magnitude than the metallic superlattices at 10 K and 88 kOe [Fig. 5 (b)], with no indication of saturation. If the spins are canted/frustrated due to competing AF/FM interactions,[23] application of a magnetic field would better align these spins, giving rise to a larger negative MR at high fields in the insulating superlattices. Furthermore, for the $n$=5 sample, the MR is *positive* at low fields and changes sign at higher fields.**Error! Bookmark not defined.** Positive MR has also been observed in multilayers with complex magnetic structure, with coexisting FM and AF couplings.[24] Further evidence for the proximity of AF/FM regions is the emergence of magnetic pinning with increasing $n$ as evidenced by an increase in $H_c$ from 20 Oe for the $n$=1 superlattice to 1100 Oe for a $n$=5 sample [Fig. 3].

Our results are consistent in some aspects with the theoretical work of Lin et al.,[5] where the long-range Coulomb interaction was taken into account with a constant dielectric function. However, in the vicinity of a metal-insulator transition from the insulating side, the dielectric function in reciprocal space $\varepsilon(q)$ tends to diverge as $1/q^2$, leading to a complete screening of the long-range Coulomb interaction in the metallic state. This divergence is cut off for $q < q_c \sim 1/\xi_L$ (*i.e.* $\varepsilon(q)$ stops increasing[25] below $q_c$), and the screening is ineffective for length scales > $\xi_L(n)$. Since $L_{TF}$ depends on $\varepsilon$, it also depends on $\xi_L(n)$. A realistic calculation for a superlattice needs to take this dependence of $L_{TF}$ on $n$ into account. Secondly, while the local hole density may be an indicator of the phase obtained in a given layer, the interaction between order parameters in neighboring layers are critical for determining properties such as conductivity and magnetization of these states.

In conclusion, we have established a that the metal-insulator transition is accompanied by a strongly modulated FM order in insulating $(LMO)_{2n}/(SMO)_n$ superlattices. The 'interfacial insulator' that emerges obeys variable-range hopping at low temperatures, consistent with a finite DOS at $E_F$. Coexisting FM and AF regions within the superlattices may give rise to the canting/magnetic disorder that localizes these states. A metal is obtained when these interfaces are brought into close proximity, consistent with a Mott-transition. The proximity of the phases in the superlattices to a metal-insulator transition, and the intimate coexistence of phases with different order parameters within the high-$n$ superlattices may give rise to a large susceptibility to external electric and magnetic fields, and holds the promise of engineering new types of mixed-phase and interfacial materials.

Work at Argonne is supported by the U.S. Department of Energy, Office of Basic Energy Sciences under contract no. DE-AC02-06CH11357. Work at the University of Illinois at Urbana-Champaign was supported by the U.S. Department of Energy, Division of Materials Science under award  no. DEFG02-91-ER45439 through the Frederick Seitz Materials Research Laboratory. This work has benefited from the use of the Lujan Neutron Scattering Center at LANSCE, which is funded the Department of Energy's Office of Basic Energy Sciences. Los Alamos National Laboratory is operated by Los Alamos National Security LLC under DOE Contract DE-AC52-06NA25396.

* email: anand@anl.gov

**Figure Captions:**

FIG. 1 (a) X-ray reflectivity and diffraction for $n$ = 1 to 5 (b) STEM z-contrast images from an $n$ = 4 superlattice showing $(LMO)_8/(SMO)_4/(LMO)_8$ regions. Rows of atoms are shown on the right as a guide to the eye.

FIG. 2.  Resistivity of $La_{0.67}Sr_{0.33}MnO_3$  random alloy film, and corresponding $(SrMnO_3)_n/(LaMnO_3)_{2n}$ superlattices, $1 \leq n \leq 5$. The inset shows the  resistivity of a pure $LaMnO_3$ thin film and  $(SrMnO_3)_3/(LaMnO_3)_1$ superlattice for reference.

Fig. 3 (a) Magnetization vs H for $n$ = 1 to 5 at $T$ = 10 K (b) Evolution of the saturation magnetization and coercivity at 10 K with increasing $n$

FIG. 4. (a) Polarized neutron reflectivity measurements of an $n$ = 5  superlattice at 300 K and (inset) the x-ray reflectivity data for the same sample. (b) PNR at 10K in a field of 5.5 kOe ($T_C$ = 180 K). SLD (y-axis label on inset) stands for scattering length density and is a measure of the neutron scattering potential. The inset shows the inferred magnetic structure from the best fit, where the shaded region is the extent of LMO in one superlattice period. (c)  PNR measurements for the $n$ = 3 sample, and (inset) the inferred magnetic structure from a calculation that best reproduced the data.

Fig. 5. (a) Fit of the low temperature data for the $n$ = 3 - 5 superlattices to the form for Mott's variable range hopping in 3D: $\rho = \rho_o \exp(T_O /T)^{1/4}$   (b) MR at 10 K. The insulating superlattices

($n \geq 3$) show a  higher value of magnetoresistance (> 44.5% at 88 kOe) than the metallic samples

($n \leq 2$) (< 6.3% at 88 kOe).

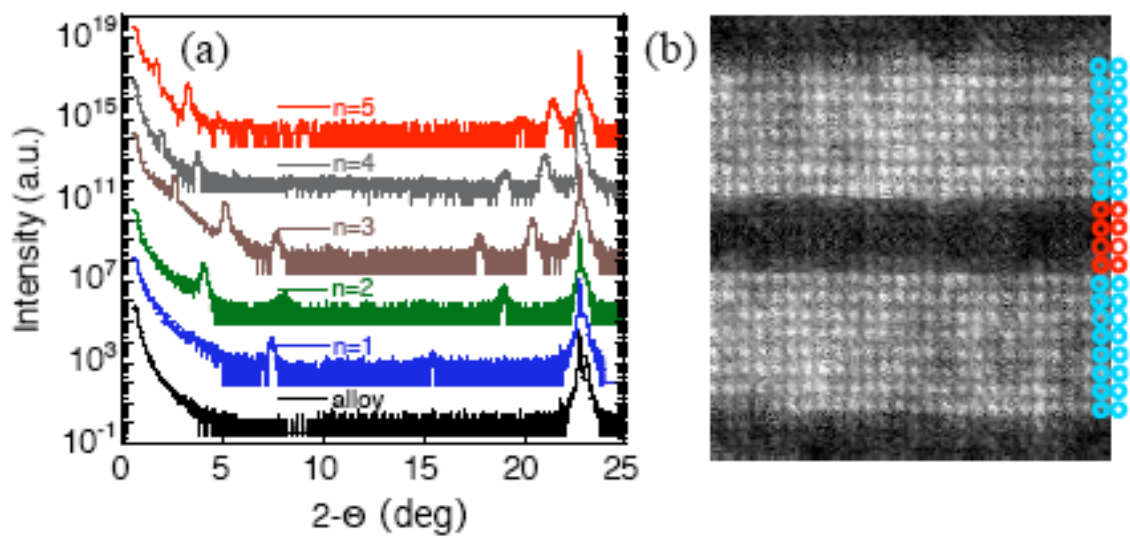

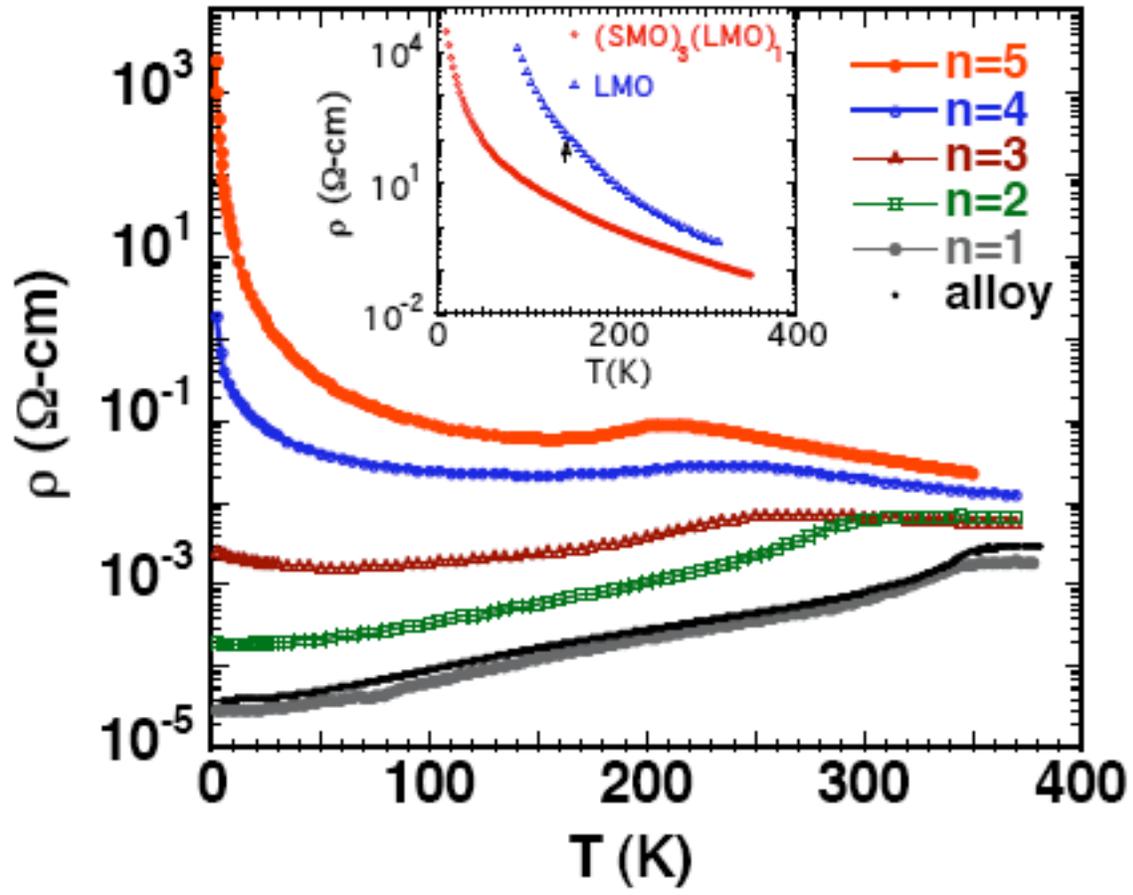

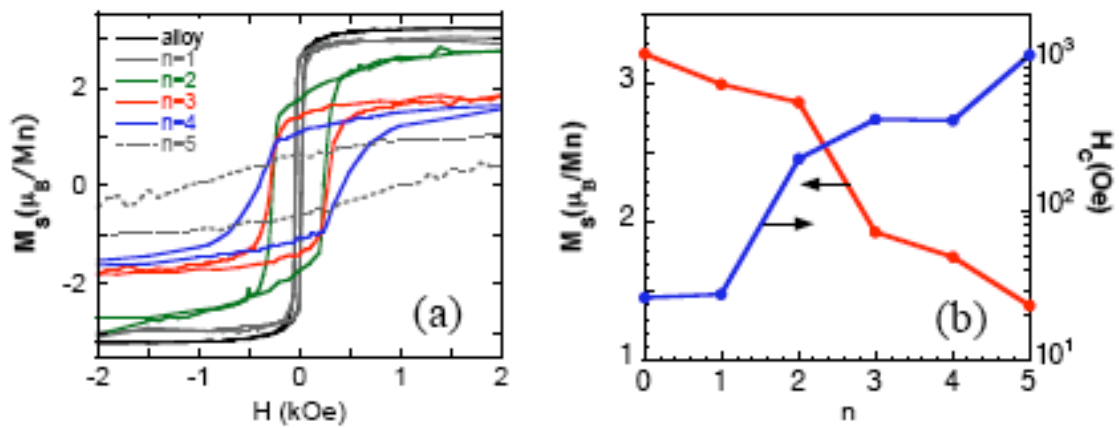

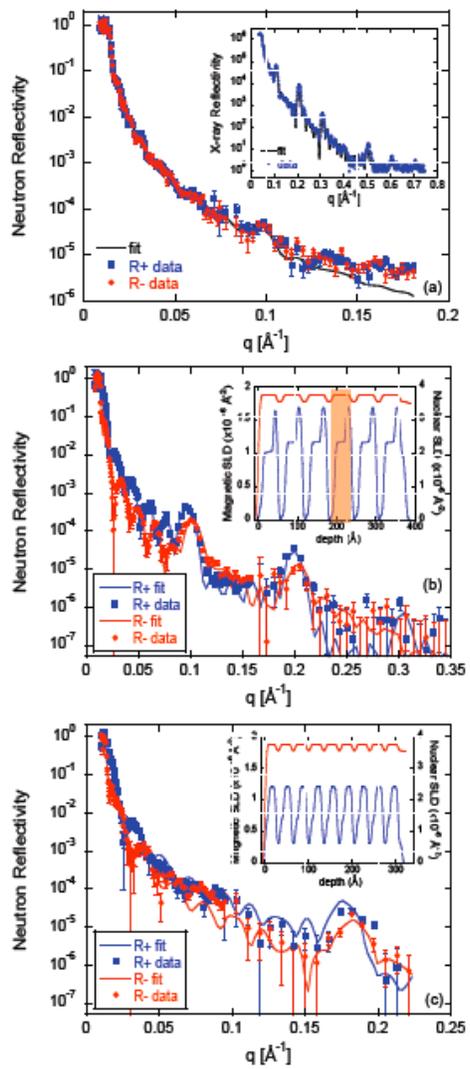

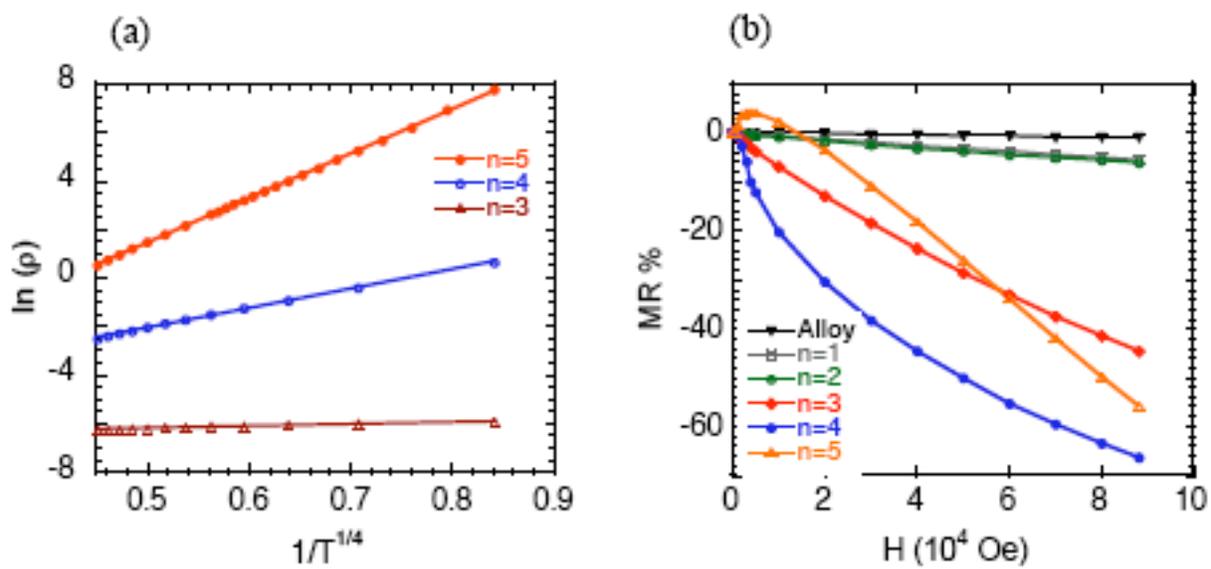

References.